\documentclass[12pt]{article}
\usepackage{latexsym}
\usepackage{here}
\usepackage[dvipdfmx]{graphicx}

\usepackage{hyperref}
\usepackage{amsmath,amssymb}
\usepackage{url}
\usepackage{graphicx,color}
\usepackage{multirow,array}

\newcommand{\PA}[1]{\biggl(#1 \biggr)}

\newcommand{\bbb}[1]{\mathbb{#1}}

\newcommand{\EX}[1]{\exp{\biggl(#1 \biggr)}}

\newcommand{\EEEX}[1]{\exp{\left[\vbox to 21pt{} #1 \right]}}
\newcommand{\pa}[1]{\left(#1 \right)}
\newcommand{\PPA}[1]{\left(\vbox to 21pt{} #1 \right)}
\newcommand{\BR}[1]{\Biggl[#1 \Biggr]}

\newcommand{\ca}[1]{\mathcal{#1}}

\newcommand{\abs}[1]{\left|#1\right|}

\newcommand{\s}[1]{\sqrt{#1}}

\newcommand{\f}[1]{\frac{#1}}

\newcommand{\bea}{\begin{eqnarray}}
\newcommand{\eea}{\end{eqnarray}}

\def\m{{\mu}}
 
 \def\n{{\nu}}
 
 \def\d{{\delta}}

 \def\a{{\alpha}}
 \def\T{{\Theta}}

 \def\b{{\beta}}

\def\ii{{\mathrm{i}}}

\def\m{{\mu}}
 
 \def\n{{\nu}}
 
 \def\d{{\delta}}

 \def\a{{\alpha}}
 \def\T{{\Theta}}
 \def\frac#1#2{{#1\over #2}}

 \def\s{\sqrt}
 
 \def\b{{\beta}}

\def\be{\begin{equation}}
\def\ee{\end{equation}}
\def\ba{\begin{eqnarray}}
\def\ea{\end{eqnarray}}
\numberwithin{equation}{section}

 \def\f {\frac}
 
 \def\ap{\alpha}

 \def\no{\nonumber \\}
\def\nn{\nonumber \\}
 \def\la{\langle}
 \def\lb{\rangle}

\begin{document}

\begin{titlepage}
\thispagestyle{empty}

\begin{flushright}
YITP-17-30
\\
IPMU17-0046
\\
\end{flushright}

\bigskip

\begin{center}
\noindent{{ \textbf{Out-of-Time-Ordered Correlators in  $(T^2)^n/\mathbb{Z}_n$}}}\\
\vspace{2cm}
Pawel Caputa$^{a}$, Yuya Kusuki$^{a}$, Tadashi Takayanagi$^{a,b}$ and Kento Watanabe$^{a}$
\vspace{1cm}

{\it
$^{a}$Center for Gravitational Physics, Yukawa Institute for Theoretical Physics, \\
Kyoto University, 
Kyoto 606-8502, Japan\\
$^{b}$Kavli Institute for the Physics and Mathematics of the Universe,\\
University of Tokyo, Kashiwa, Chiba 277-8582, Japan\\
}

\vskip 2em
\end{center}

\begin{abstract}
In this note we continue analysing the non-equilibrium dynamics in the $(T^2)^n/\mathbb{Z}_n$ orbifold conformal field theory. We compute the out-of-time-ordered four-point correlators with twist operators. For rational $\eta \ (=p/p')$ which is the square of the compactification radius, we find that the correlators approach non-trivial constants at late time. For $n=2$ they are expressed in terms of the modular matrices and for higher $n$ orbifolds are functions of $pp'$ and $n$. For irrational $\eta$, we find a new polynomial decay of the correlators that is a signature of an intermediate regime between rational and chaotic models.  
\end{abstract}
\end{titlepage}

\newpage

\section{Introduction}
Conformal field theories (CFTs) in two dimensions play a very important role in modern physics. From describing critical points of quantum Hamiltonians to AdS/CFT, they provide a powerful framework to understand physical phenomena in interacting many-body systems like e.g. thermalization \cite{Qa} or entanglement \cite{Ea,Eb}. Nevertheless, even though the conformal symmetry drastically constraints these theories and many things about 2d CFTs are known \cite{DiF}, we are neither able to classify  even rational CFTs nor a pinpoint of 2d CFTs which have classical holographic duals.

Recently, a progress on the later issue has been obtained in the context of black holes in the AdS/CFT correspondence. More precisely, by studying small perturbations to black holes, authors of \cite{Shenker:2014cwa} proposed that a certain out-of-time-ordered correlators (OTOC) could be used to diagnose the butterfly effect\footnote{This was further formalized by Kitaev \cite{Kitaev} that connected OTOC with older semiclassical diagnose of quantum chaos in the form of the expectation value of the square of the commutator \cite{LO}.}\footnote{See also \cite{Caputa:2015waa} for the CFT computation.}. Subsequently, in \cite{Maldacena:2015waa}, it was argued that theories with Einstein gravity dual should exhibit the maximal Lyapunov exponent as measured by the OTOC. If correct, this important progress provides a necessary condition for a CFT to behave holographically. It is then important to test OTOC in known models and understand why they are good measures of quantum chaos and the smoking gun of holography.

The out-of-time-ordered thermal correlator can be defined as the inner product between two states: $W(t)V\left|0\right>_\beta$ and $VW(t)\left|0\right>_\beta$ where the "generic" operators $W$ and $V$ are separated in space by $x$ and in Lorentzian time $t$, and $\beta$ is an inverse temperature\footnote{The definition that we take here is the original that appeared in the context of black holes \cite{Shenker:2014cwa}. Clearly, OTOC can be generalized to zero temperature and arbitrary operators.}. The OTOC has then a form of the normalized four-point correlator
\be
C_\beta(x,t)\equiv\frac{\langle V^\dagger W^\dagger(t)V W(t)\rangle_\beta}{\langle V^\dagger V\rangle_\beta\langle W^\dagger W\rangle_\beta}.
\ee
As for any Lorentzian correlator in quantum field theory, the out-of-time ordered correlator can be obtained from the Euclidean four point function by analytical continuation \cite{Roberts:2014ifa}. More precisely, starting from $\langle V^\dagger_1V_2 W^\dagger_3 W_4\rangle_\beta$ (where $O_i\equiv O(z_i,\bar{z}_i))$ we "order" the operators along the imaginary time: $\epsilon_1<\epsilon_3<\epsilon_2<\epsilon_4$, and then analytically continue to the real time such that the insertion points of the operators on the thermal cylinder are 
\bea
z_1&=&e^{\frac{2\pi}{\beta}(t+ \ii \epsilon_1)},\qquad \bar{z}_1=e^{-\frac{2\pi}{\beta}(t+ \ii \epsilon_1)},\nn
z_2&=&e^{\frac{2\pi}{\beta}(t+\ii \epsilon_2)},\qquad \bar{z}_2=e^{-\frac{2\pi}{\beta}(t+\ii \epsilon_2)},\nn
z_3&=&e^{\frac{2\pi}{\beta}(x+\ii \epsilon_3)},\qquad \bar{z}_3=e^{\frac{2\pi}{\beta}(x- \ii \epsilon_3)}, \nn
z_4&=&e^{\frac{2\pi}{\beta}(x+\ii \epsilon_4)},\qquad \bar{z}_4=e^{\frac{2\pi}{\beta}(x-\ii \epsilon_4)}.
\eea
From these points, we form the standard conformal cross-ratios $(z,\bar{z})$, (see App. \ref{CROTO}) such that, as time progresses, the chiral conformal cross-ratio crosses a branch-cut
\be
1-z \to e^{-2\pi \ii}(1-z), \label{OTOCont}
\ee
whereas the anti-chiral remains approximately zero $\bar{z}\simeq 0$. We refer to this continuation as the {\it OTO continuation}. In order to extract the chaotic features of the OTOC, we focus on the "late time" behaviour of the continued correlator by taking
\be
z,\bar{z}\to0 ,\qquad \text{with} \qquad \bar{z}/z-\text{fixed}. \label{LateTimeLim}
\ee
The explicit dependence on the inverse temperature $\beta$ enters through the late time expressions
\bea
z\simeq -e^{-\frac{2\pi (t-x)}{\beta}}\epsilon^*_{12}\epsilon_{34},\qquad
\bar{z}\simeq -e^{-\frac{2\pi (t+x)}{\beta}}\epsilon^*_{12}\epsilon_{34},\label{CRbeta}
\eea
where $\epsilon_{ij}=\ii\left(e^{\frac{2\pi \ii}{\beta}\epsilon_i}-e^{\frac{2\pi \ii}{\beta}\epsilon_j}\right)$ and $x$ is the separation between operators that keeps the ratio $\bar{z}/z$ fixed.\\
From the explicit computations of the OTOC (see an early review in e.g. \cite{Perlmutter:2016pkf}), it is believed that in chaotic models with a large separation of scales between $t=\beta$ and $t_*\sim\beta\log c$ (the so-called scrambling time), the early exponential decay of the OTOC is governed by bounded the Lyapunov exponent $\lambda_L\le 2\pi/\beta$ which is saturated by holographic models with "maximal chaos". Then, after the scrambling time, the OTOC in chaotic  systems decay to zero exponentially. Nevertheless, the complete classification of the OTOC evolutions in quantum many-body systems is now a very active area of research that will further elucidate their importance.

From the very beginning the OTOCs attracted a lot of attention and found important applications in various physical scenarios (see e.g. \cite{Roberts:2014ifa,Perlmutter:2016pkf,OTOs,Caputa:2016tgt,Gu:2016hoy}). In this work we use them as a tool to classify CFTs from the point of view of quantum information spreading. More precisely, we focus on the features of OTOCs in a family of CFTs defined by the sigma model whose target space is the cyclic orbifold $(T^2)^n/\mathbb{Z}_n$. In this setup, in addition to "generic" primary operators, we have another "natural" candidates for the operators $W$ and $V$, namely the twist operators $\sigma_n$. Since the twist operators are directly used to compute entanglement measures (and their evolution) in 2d CFTs, on may expect that they should also be good operators to probe quantum chaos. In this work, we employ them in the computation of the OTOC and make a modest progress in sharpening this intition.

Last but note least, in \cite{Caputa:2017tju}, we studied the time evolution of Renyi entanglement entropy for a locally excited state created by a twist operator in this orbifold CFT. We showed that depending on the square of the compactification radius $R$ denoted as $\eta = R^2$, Renyi entropies either increase by the amount of the quantum dimension for rational compactification parameters $\eta = p/p'$, following the general rules found in \cite{QD}, which also looks similar to the free CFTs \cite{localEE}. For irrational compactification parameters $\eta \neq p/p'$, interestingly it grows as a double logarithm of time $\log\log t$, which is milder than the logarithmic growth $\log t$ \cite{HEE} of the holographic CFTs. Therefore it is very interesting to understand if this new class of entanglement growth (probably characteristic for irrational CFTs) is already enough to classify these CFTs as chaotic ones from the OTOC point of view. This is our main motivation and below we shed more light on this question.

This paper is organized as follows. In section \ref{OTOInt} we present a general definition of the OTOC in our setup. Section \ref{n2O} contains results for $n=2$ orbifolds computed in two different approaches and in section \ref{nHO} we present a direct computation for higher $n$ orbifolds. Finally, in section \ref{Conclusions} we summarize our findings and conclude. 

\section{OTOC in $(T^2)^n/\mathbb{Z}_n$}\label{OTOInt}
In this work, we focus on the cyclic orbifold CFT  $(T^2)^n/\mathbb{Z}_n$, where $T^2=S^1\times S^1$ is the $c=2$ CFT defined by two scalar bosons which are compactified on the same radius $R$.
We consider four-point OTOCs where we choose as the operators $W$ and $V$ the primary operator $\sigma_n$
\be
V(z,\bar{z})=W(z,\bar{z})=\sigma_n(z,\bar{z}),
\ee
where $\sigma_n$ is the twist operator for the cyclic transformation of $n$ copies of $T^2$, which has the conformal dimension $\Delta_n=\bar{\Delta}_n=\frac{1}{12}\left(n-\frac{1}{n}\right)$. Our main aim will be to explore how the OTOCs evolve with time depending on the compactification parameter $\eta=R^2$. In general, we will distinguish between rational $\eta=p/p'$ and irrational $\eta\neq p/p'$ or rational and irrational CFTs respectively.

According to the general arguments in \cite{Caputa:2016tgt,Gu:2016hoy}, in rational CFTs, the Lyapunov exponent is zero and the OTOC approaches to a constant at late time. The constant is equal to the $(0,0)$ element of the monodromy matrix and for our twist operators is expected to be 
\be
C_\beta(x,t)\to M_{00}=\frac{S^*_{\sigma_n\sigma_n}}{S_{00}d^2_{\sigma_n} },
\ee
where the modular $S$-matrix components (complex conjugate) and the quantum dimensions correspond to the twist operators. Even though there exists a general (formal) expression for the $S^*_{\sigma_n\sigma_n}$ in orbifold theories \cite{Orbifold}, we have not been able to extract the explicit data to match our setup \footnote{More precisely, the formulas in \cite{Orbifold} are given for arbitrary orbifolds and are expressed in terms of formal mathematical objects. We were neither able to evaluate them nor find their explicit forms in the literature for our particular setup.} and we proceeded with direct computation instead.\\
In irrational CFTs the general characteristics of the OTOCs are not known and our model is an analytic setup where this computation can be performed for the first time.

Let us now briefly review the main correlator that we will need in this work (since we only review the necessary minimum, for more details see \cite{CCT,Caputa:2017tju}). The Euclidean four-point function that we will use has a general form 
\ba
\la \sigma_n(z_1,\bar{z}_1)\bar{\sigma}_n(z_2,\bar{z}_2)\sigma_n(z_3,\bar{z}_3)
\bar{\sigma}_n(z_4,\bar{z}_4)\lb
=|z_{12}z_{34}|^{-4\Delta_n}|1-z|^{-4\Delta_n}F_n(z,\bar{z}), \label{fnd}
\ea
where the conformal cross-ratios are defined in the usual way as $z=\frac{z_{12}z_{34}}{z_{13}z_{24}}$ and $\bar{z}=\frac{\bar{z}_{12}\bar{z}_{34}}{\bar{z}_{13}\bar{z}_{24}}$, and the main information about the theory is encoded in function $F_n(z,\bar{z})$.\\
For the cyclic orbifold CFT $(T^2)^n/\bbb{Z}_n$ the function $F_n(z,\bar{z})$ is expressed as follows \cite{CCT}
\ba
F_n(z,\bar{z})\equiv \f{\T^2(0|\text{T})}{\prod_{k=1}^{n-1}f_{k/n}(z)f_{k/n}(\bar{z})}. \label{xxq}
\ea
In the above, the Riemann-Siegel theta function is expressed in terms of matrix matrix $\text{T}$ that is itself a function of $\tau$ and $\bar{\tau}$ (see \cite{CCT} for details) in a way that the function reads
\begin{equation}\label{RST}
\T(0|\text{T})\equiv\sum_{\mathbf{m},\mathbf{n} \in \mathbb{Z}^{n-1}} e^{\f{\pi \ii }{2}\PA{\mathbf{n}\sqrt{\eta}+\f{\mathbf{m}}{\sqrt{\eta}}} \cdot \tau \cdot \PA{\mathbf{n}\sqrt{\eta}+\f{\mathbf{m}}{\sqrt{\eta}}}} e^{-\f{\pi \ii  }{2}\PA{\mathbf{n}\sqrt{\eta}-\f{\mathbf{m}}{\sqrt{\eta}}} \cdot \bar{\tau} \cdot \PA{\mathbf{n}\sqrt{\eta}-\f{\mathbf{m}}{\sqrt{\eta}}}},
\end{equation}
where $\mathbf{n}$ and $\mathbf{m}$ are $(n-1)$-dimensional vectors and the expression for the period matrix is
\begin{equation}
\label{eq:period}
(\tau)_{ij} = \f{2}{n} \sum_{k=1}^{n-1}
\bigg[\, \ii \, \f{f_{k/n}(1-z)}{f_{k/n}(z)} \,\bigg]\sin\left(\f{\pi k}{n}\right)
\cos\left(\f{2\pi k(i-j)}{n}\right)\,.
\end{equation}
and analogously for $\bar{\tau}$. Finally, function in the denominator is the hypergeometric function
\ba
\label{eq:Ikn}
f_{k/n}(z,\bar{z})=\,_2F_1(k/n,1-k/n,1,z),
\ea

In the following, we focus on the OTOCs that are written in terms of the above data as
\be
C_\beta(x,t)=|1-z|^{-4\Delta_n}F_n(z,\bar{z}),
\ee
with the cross-ratios defined in App. \ref{CROTO}. In order to understand the general implications of this correlator for OTOC, it will be instructive to consider the $n=2$ first and we begin with this case.

Before we proceed, let us point out that there are several reasons to consider OTOC with twist operators. Most importantly, using the replica approach, they are naturally linked to the quantum information measures like entanglement entropy and mutual information so we can compare the information from all these quantities with OTOC. Moreover, for $n=2$ the four-point correlators are directly expressed by the partition function of the seed theory. Therefore, in the light of quantum chaos measures, these correlators should be useful in confronting the information from the spectral from factor \footnote{We would like to thank Gabor Sarosi for explaining the details of the spectral form factors \cite{FF} and preliminary confirmation of the consistency with our results.}.
\section{The $n=2$ orbifold}\label{n2O}
The $n=2$ orbifold is very special and universal in terms of the four-point function of the twist operators. The $\Theta$ function becomes 
\begin{equation}
\label{eq:theta}
\T(0|\text{T})=\sum_{\m,\n \in \mathbb{Z}} \EX{\f{\pi \ii \tau }{2}\PA{\n\sqrt{\eta}+\f{\m}{\sqrt{\eta}}}^2} \EX{-\f{\pi \ii \bar{\tau} }{2}\PA{\n\sqrt{\eta}-\f{\m}{\sqrt{\eta}}}^2},
\end{equation}
and in fact the correlator can be directly written in terms of the partition function of the seed theory as\footnote{In general, in the $n=2$ orbifolds the four-point function of the twist operators is proportional to the partition function of the seed theory which in our case is the square of the compactified boson partition function.}
\be
F_2(z,\bar{z})=\f{\T^2(0|\text{T})}{f_{1/2}(z) f_{1/2}(\bar{z})}=2^{-4/3}|z(1-z)|^{1/3}Z^2_{\eta}(\tau,\bar{\tau}).
\ee
The partition function of the compactified boson is 
\be
Z_{\eta}(\tau,\bar{\tau})=\frac{\Theta(0|\text{T})}{\abs{\eta(\tau)}^2}=\frac{1}{\abs{\eta(\tau)}^2 }\sum_{\mu,\nu\in\mathbb{Z}}q^{\frac{\eta}{4}\left(\mu+\frac{\nu}{\eta}\right)^2}\bar{q}^{\frac{\eta}{4}\left(\mu-\frac{\nu}{\eta}\right)^2},\label{Partition}
\ee
where $\eta(\tau)$ is the Dedekind eta function and $q=e^{2\pi \ii\tau}$ while $\bar{q}=e^{-2\pi \ii \bar{\tau}}$. The modular parameter is related to the cross-ratios as
\be
\tau=\ii\frac{f_{1/2}(1-z)}{f_{1/2}(z)}=\ii\f{K(1-z)}{K(z)},\label{MODpar}
\ee
with $K(z)$ is the complete elliptic integral of the first kind.

In order to obtain the out-of-time-ordering we first have to perform the analytic continuation \eqref{OTOCont} in our expression and then take the appropriate limit of the vanishing cross-ratios. To do that, we first have to translate the analytic continuation to the operation on the modular parameters. For that, we will need the fact that under the OTO continuation $(1-z)\to e^{-2\pi \ii}(1-z)$ the hypergeometric functions undergo a well known monodromy \cite{Erdelyi} 
\be
f_{1/2}(z)\to f_{1/2}(z)+2\ii f_{1/2}(1-z),\qquad f_{1/2}(1-z)\to f_{1/2}(1-z).\label{OTOf12}
\ee
This way, using the \eqref{MODpar}, we can derive that this continuation corresponds to the modular $S\bar{T}^{2}S$ transformation
\be
\tau\to \frac{\tau}{1+2\tau}.\label{STTS}
\ee
After this transformation we take the $z,\bar{z}\to 0$ which translates to $\tau\to \ii\infty$ , $\bar{\tau}\to-\ii\infty$.

Up to this point, the discussion has been general but now we will distinguish two computations for rational and irrational $\eta$. In the rational case, we can use the power of RCFTs and we begin with that. Next, we confirm that a direct computation leads to the same result and proceed with direct approach to irrational CFTs.

\subsection{Rational $\eta=p/p'$}
Here we compute the OTOC by first using the RCFT techniques and then by direct evaluation of the Riemann-Siegel $\Theta$ function with the OTO continuation (\ref{OTOCont}) in the late time limit (\ref{LateTimeLim}). The second approach will turn out to be more powerful and will be applied for irrational compactification parameters $\eta$ and higher $n$ orbifolds.

\subsubsection{RCFT approach}
For the rational compactification parameter $\eta=p/p'$ with integers $p$ and $p'$, we can write the partition function as a finite sum of characters \cite{DiF}
\be
Z_{p/p'}(\tau,\bar{\tau})=\sum^{N-1}_{\lambda=0}K_{\lambda}(\tau)K_{\omega_0\lambda}(\bar{\tau}),
\ee
where the characters are defined as
\be
K_{\lambda}(\tau)=\frac{1}{\eta(\tau)}\sum_{\nu\in\mathbb{Z}}q^{\frac{(\nu N+\lambda)^2}{2N}},\label{charK}
\ee
and $N=2pp'$, $\omega_0=p'r_0+ps_0$ mod $N$ and $(r_0,s_0)$ is a unique pair in the range $1\le r_0\le p-1$, $1\le s_0\le p'-1$, $ps<p'r$ satisfying $p'r_0-ps_0=1$ and $\omega^2_0=1$ mod $2N$.\\

Next, we apply the $S\bar{T}^2S$ transformation to $\tau$ which yields the transformed partition function
\be
Z_{p/p'}\left(\frac{\tau}{1+2\tau},\bar{\tau}\right)=\sum^{N-1}_{\lambda=0}\sum^{N-1}_{\nu=0}\left(S\bar{T}^{2}S\right)_{\lambda\nu}K_\nu(\tau)K_{\omega_0\lambda}(\bar{\tau}),
\ee
with the explicit transformation matrix (using formulas from \cite{DiF})
\be
\left(S\bar{T}^{2}S\right)_{\lambda\nu}=\frac{1}{N}e^{\frac{\pi \ii}{6}}\sum^{N-1}_{\mu=0}e^{2\pi \ii\frac{\mu\left(\lambda+\nu-\mu\right)}{N}}.
\ee
Finally, once we expand the characters for vanishing $q$ and $\bar{q}$, the dominant contribution comes from the vacuum characters and we have   
\be
Z_{p/p'}\left(\frac{\tau}{1+2\tau},\bar{\tau}\right)\to \left(S\bar{T}^{2}S\right)_{00}q^{-\f{1}{24}}\bar{q}^{-\f{1}{24}}=\left(S\bar{T}^{2}S\right)_{00}2^{2/3}|z|^{-1/6}.
\ee
where in the last step we used the relation between $\tau$ and $z$ in the limit of $z\to 0$ 
\be
\tau=\ii\frac{K(1-z)}{K(z)}\simeq \frac{\ii}{\pi}\log\left(\frac{16}{z}\right)+O(z),
\ee
and similarly for $\bar{q}$ and $\bar{z}$.\\
This way, putting all the terms together we conclude that for $n=2$ the OTOCs exponentially approach to a constant
\be
C_\beta(x,t)\to e^{4\pi \ii \Delta_2}e^{-\frac{\pi \ii}{3}}\left(S\bar{T}^{2}S\right)^2_{00}.\label{OurRes}
\ee 

It is instructive to be slightly more explicit and unpack this formula. For that, let us compute the $(0,0)$ element of the transformation matrix explicitly. Even though the expression appears as a simply-looking sum, it is by no-means trivial. In fact the sum has an old history in mathematics. Namely, Gauss proved that \cite{BerndtEvans1981,BerndtEvansWilliams1998}  
\begin{equation}
\sum^{N-1}_{\m=0}e^{ -\f{2 \pi \ii\m^2}{N}}=\left\{
    \begin{array}{ll}
      \sqrt{N}, &\text{if } N\equiv 1\, (\text{mod }4),\\
	0,  &\text{if } N\equiv 2\, (\text{mod }4),\\
      -\ii \sqrt{N}, &\text{if } N\equiv 3\, (\text{mod }4),\\
     (1-\ii) \sqrt{N}, &\text{if } N\equiv 0\, (\text{mod }4),
    \end{array}
  \right.\\
\end{equation}
where $N$ is any natural number. Then using this identity, we can derive
\begin{equation}
(S \bar{T}^{2} S)_{00}=\left\{
	\begin{array}{ll}
      \f{1}{\sqrt{pp'}} e^{-\f{\pi \ii}{12}}, &\text{if } pp' \in \text{even},\\
      0, &\text{if } pp' \in \text{odd}.
    \end{array}
  \right.\\
\end{equation}
As we can see, the constant that OTOC approaches can actually be zero so let us also take the sub-leading terms into account. 

For the sub-leading contributions, it becomes important to specify if we are working in the large or small temperature limit. Namely, in our formulas, if we are interested in terms of order $q^{1/2N}$, for small temperatures we should also take into account terms that scale as $\bar{q}^{1/2N}$. This issue of limits is of course specified by the ratio $\bar{z}/z\sim e^{-\frac{4\pi x}{\beta}}$ and we will assume that $x/\beta$ is large i.e. high temperatures. 

This way, in the high temperature limit, there are two sub-leading terms of order $q^{\f{1}{2N}}$ and we have
\be
Z_{p/p'}\left(\frac{\tau}{1+2\tau},\bar{\tau}\right)\to \left[\left(S\bar{T}^{2}S\right)_{00}+\left(\left(S\bar{T}^{2}S\right)_{01}+\left(S\bar{T}^{2}S\right)_{0, N-1}\right)q^{\f{1}{2N}}\right]q^{-\f{1}{24}}\bar{q}^{-\f{1}{24}}.
\ee
For odd $pp'$, where the first component vanishes, we can use the generalized Landsberg-Schaar identity \cite{BerndtEvans1981} to show that
\begin{equation}
(S \bar{T}^2 S)_{01}=(S \bar{T}^2 S)_{0, N-1}=\f{1}{\sqrt{pp'}} e^{-\f{\pi \ii}{12}} e^{\frac{\ii\pi}{4pp'}}.
\end{equation}
Summarizing, we have shown that for the rational compactification parameter $\eta=p/p'$, the OTOC approaches exponentially to a constant $\eqref{OurRes}$. For odd $pp'$, the $(0,0)$ matrix element vanishes and the OTOC decays exponentially to zero. In terms of the $pp'$ we can write it then as
\begin{equation}\label{OTORCFT}
\begin{aligned}
C_\beta(x,t)\to e^{4\pi i \Delta_2}\left\{
	\begin{array}{ll}
      -\f{\ii}{pp'} , &\text{if } pp' \in \text{even},\\
      -\f{4\ii}{pp'}\PA{\f{-\ii\epsilon^*_{12}\epsilon_{34} }{16}}^{\f{1}{pp'}}e^{-\frac{2\pi}{pp'\beta}(t-x)}, &\text{if } pp' \in \text{odd}.
    \end{array}
  \right.\\
\end{aligned}
\end{equation}  
This our main result for the rational compactification parameter that we obtained using the fact that the theory is a RCFT and its partition function can be written in terms of finite number of characters. Below we will generalize this derivation so that it can be also applied to irrational compactification parameters and higher $n$ orbifolds. 

\subsubsection{Direct approach}
Another way to extract the OTOC is by performing the continuation \eqref{OTOCont} and take the limit $(z,\bar{z}) \equiv (16 e^{-\f{\pi}{\d}}, 16 e^{-\f{\pi}{\d^{*}}}) \to (0,0)$ with fixed ratio $\bar{z}/z$. In the above, we introduced $\delta$ and $\d^{*}$ which are both taken to zero but with relative "speed" that depends on the temperature.\\ 
As a result, we have to evaluate \eqref{eq:theta} directly as a function of the period matrix
\begin{equation}
\label{eq:taumono}
\tau \simeq \f{1}{2-\ii \d},~~~\bar{\tau}\simeq -\f{\ii}{\d^{*}}.
\end{equation}
The dominant contribution in the limit $\d^{*} \to 0$ from the anti-holomorphic part of $\T(0|T)$ can be evaluated by the saddle point approximation, which yields the following condition
\begin{equation}
\label{eq:dominant}
\n \eta -\m=0.
\end{equation}
For $\eta=\f{p}{p'}$ with $p,p'$ two positive, coprime integers, this condition can be satisfied by $\m=p \tilde{\m}, \n = p' \tilde{\m} \ (\tilde{\m} \in \mathbb{Z}) $. Similarly, the sub-leading contribution is obtained when the following condition is satisfied
\begin{equation}
\n \eta -\m=1.
\end{equation}
Let's recall the fact that the integers $p$ and $p'$ being coprime, Bezout's lemma states that there exists a couple of integers $(x,y)$ such that
\begin{equation}
\label{eq:Bezout}
p'x-py=1,
\end{equation}
and if we find the pair $(x,y)$, all pairs satisfying (\ref{eq:Bezout}) are expressed as
\begin{equation}
(x+kp,y+kp'),~~~k \in \mathbb{Z}.
\end{equation} 
In order to make the following calculation clear, we define
\begin{equation}
p'x_{\pm}-py_{\pm}=\pm 1,
\end{equation}
where $(x_{\pm},y_{\pm})$ is one pair satisfying this equation.\\
As a result of the above arguments, we can approximate \eqref{eq:theta} expressed by OTO continued $\tau$ and $\bar{\tau}$ in \eqref{eq:taumono} as
\begin{equation}
\begin{aligned}
\T(0|\text{T}_{mono})&\simeq \sum_{\tilde{\m} \in \mathbb{Z}} e^{2 \pi \ii p p' \tilde{\m}^2 \f{1}{2-\ii \d}}
+\sum_{\a=\pm}e^{-\f{\pi}{2pp'\d^{*}}} \sum_{k \in \mathbb{Z}} e^{\f{\pi \ii \tau}{2pp'} \pa{p'x_{\a}+py_{\a}+2kpp'}^2},
\end{aligned}
\end{equation}
where we introduced a label $\text{T}_{mono}$ to distinguish from matrix $\text{T}$ before the OTO continuation.\\
Now the crucial point is that the sums in this expression can be expressed as the Jacobi theta function and using its modular properties yields
\begin{equation}
\label{eq:totyu}
\begin{aligned}
&\T(0|\text{T}_{mono})=\theta_3\PA{\f{2 p p'}{2-\ii \d},0}
+\sum_{\a=\pm}e^{-\f{\pi}{2pp'\d^{*}}} e^{\f{\pi \ii}{2pp'} \f{1}{2-\ii \d} (p'x_{\a}+py_{\a})^2}  \theta_3 \PA{\f{2pp'}{2-\ii \d},\f{p'x_{\a}+py_{\a}}{2-\ii \d}}\\
&\simeq \left\{
    \begin{array}{ll}
      \sqrt{\f{\d}{p p' \ii \d}}, &\text{if }  p p' \in \text{even},   \\
      \sqrt{\f{\d}{p p' \ii \d}}\PPA{
 	2e^{-\f{\pi}{2pp'\d^{*}}}+2e^{-\f{\pi}{2pp'\d}}}
	, &\text{if }  p p' \in \text{odd}.
    \end{array}
  \right.\\
\end{aligned}
\end{equation}
Finally, we focus on the high temperature limit, which is equivalent to
\begin{equation}
\f{\bar{z}}{z}=e^{-\f{4\pi x}{\b}} \to 0,
\end{equation}
which means $\d\gg\d^*$. In this limit, we can neglect the first term of (\ref{eq:totyu}) and after inserting to the OTOC we reproduce \eqref{OTORCFT}.\\

We verify our formulas numerically and plot the absolute value of the $F_2(t)\equiv F_2(z(t),\bar{z}(t))$ in Fig. \ref{fig:Rabs}.

{\setlength\abovecaptionskip{5pt}
\setlength\intextsep{8pt}
\begin{figure}[htbp]
 \begin{minipage}{0.5\hsize}
 \begin{center}
  \includegraphics[width=7.0cm,clip]{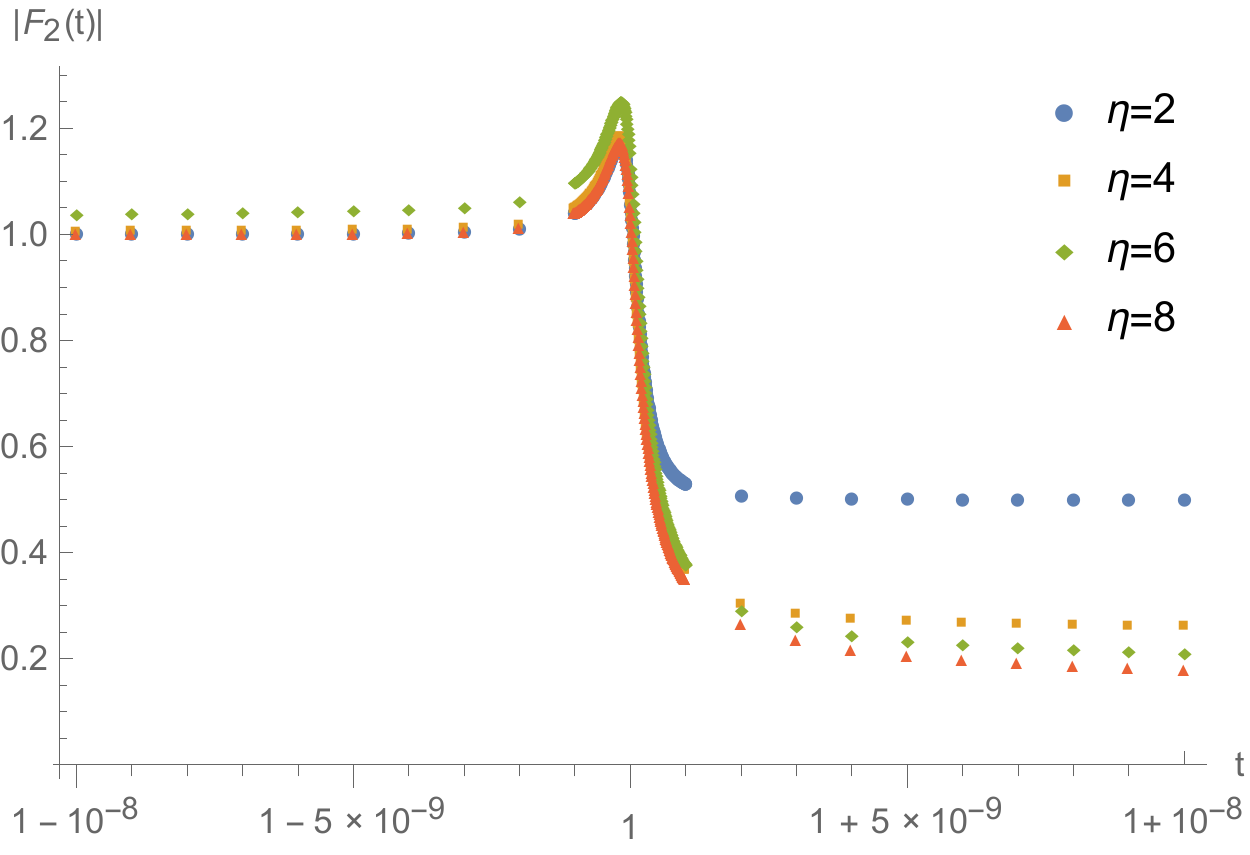}
 \end{center}
 \end{minipage}
 \begin{minipage}{0.5\hsize}
 \begin{center}
  \includegraphics[width=7.0cm,clip]{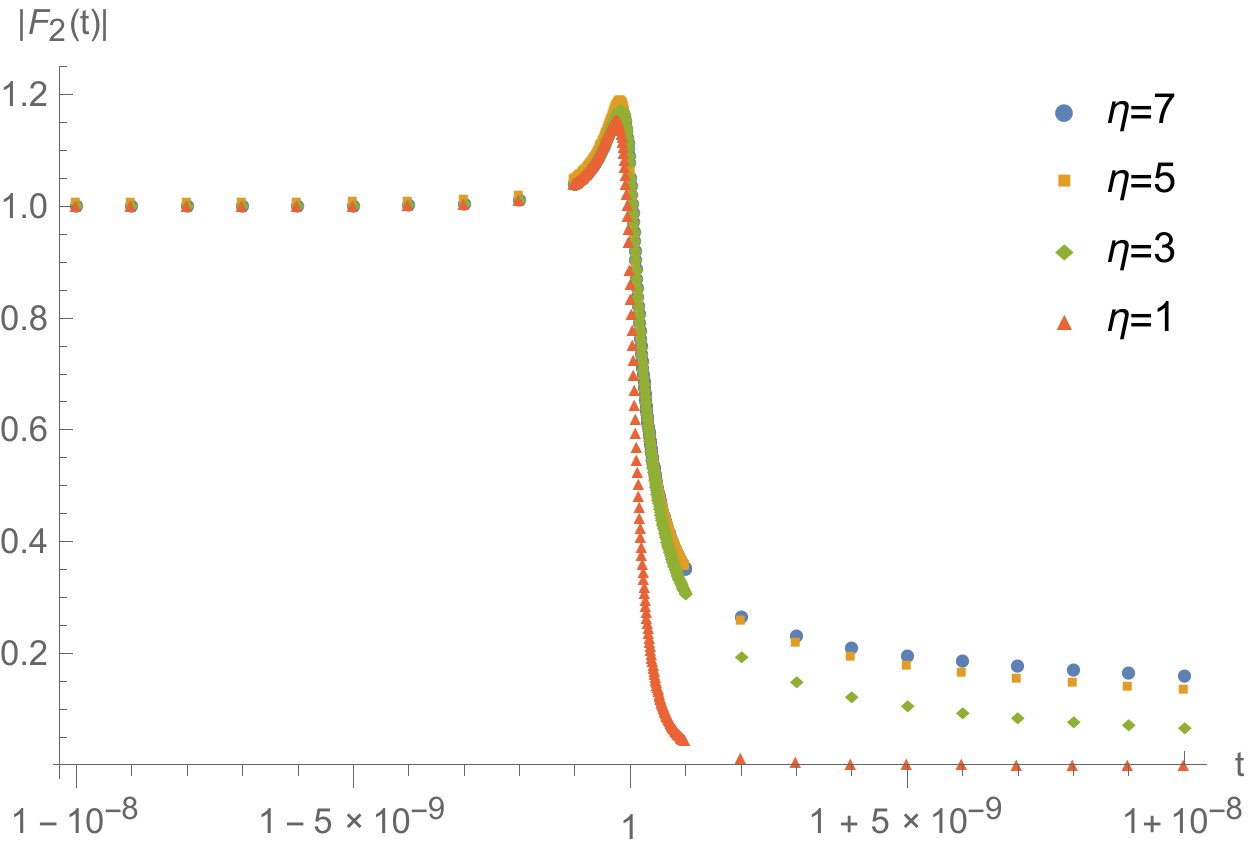}
 \end{center}
 \end{minipage}
 \caption{The plots of the absolute value $| F_{2} (t) |$ for even $\eta= 2, 4, 6, 8$ (left) and odd $\eta = 1, 3, 5, 7$(right).  We set $|t-1|<10^{-8}, x=1, \beta =0.1, 
\epsilon_1 = \epsilon_2/6 = \epsilon_3/4 = \epsilon_4/8 = 10^{-10}$.}\label{fig:Rabs}
\end{figure}
}

\subsection{Irrational $\eta\neq p/p'$}
\begin{figure}[htbp]
 \begin{center}
  \includegraphics[width=7.5cm]{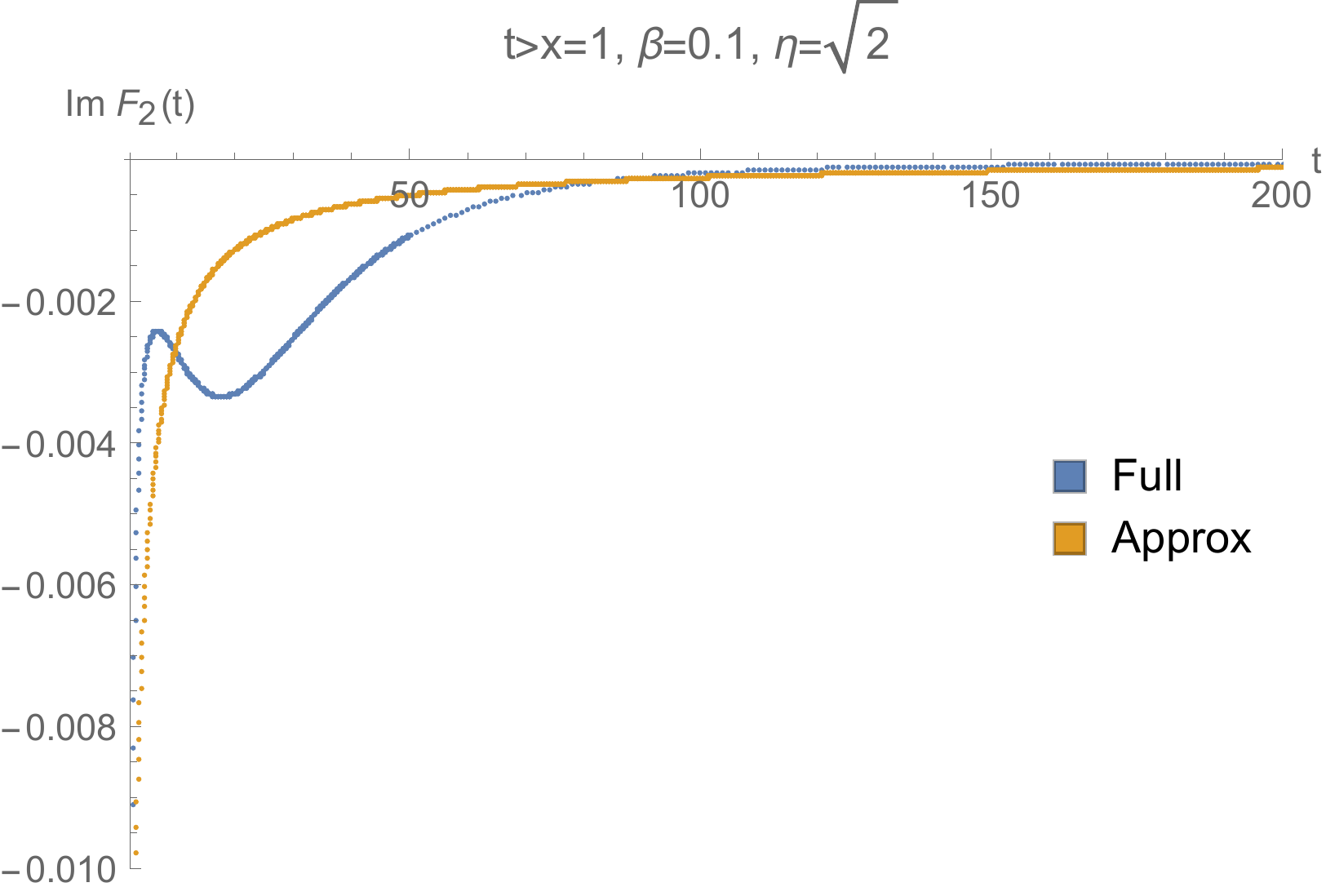}
 \end{center}
 \caption{Left : Full numerical late time plots of $\mathrm{Im} \ F_{2} (t)$ for $\eta = \sqrt{2}$ (blue) and plots of the analytically approximated formula (orange).
}\label{fig:N2sr2}
\end{figure}

For irrational compactification parameters $\eta\neq p/p'$, we can only use the direct computation which is based on extracting the limit \eqref{eq:taumono} from the $\Theta$ function.
Following the same steps as in the previous section we arrive at condition (\ref{eq:dominant}) which for $\eta\neq p/p'$ is only satisfied by $\mu=\nu=0$.  Practically, this amounts to approximating $\Theta(0|\text{T}_{mono})\simeq 1$ or, in other words, reduces the partition function to (the square of) the partition function for non-compactified boson given by the inverse of the $|\eta(\tau)|^2$.\\
The relevant function $F_2(z,\bar{z})$ that governs the four-point correlator can be approximated in this limit by
\be
F_2(z,\bar{z})\simeq \frac{1}{f_{1/2}(z)}.
\ee
Therefore, using \eqref{OTOf12}, after the OTO continuation and at late time, function $F_2(z,\bar{z})$ decays logarithmically with the holomorphic cross-ratio 
\be
F_2(z,\bar{z})\to\frac{1}{f_{1/2}(z)+2\ii f_{1/2}(1-z)}\simeq \f{\ii\pi}{2\log\left(\f{z}{16}\right)}.
\ee
Finally, inserting the explicit form of the cross-ratios \eqref{CRbeta} with Lorentzian time and inverse temperature $\beta$, gives the OTOC for irrational $\eta$
\be
C_\beta(x,t)\to e^{4\pi \ii\Delta_2}\frac{\ii\pi}{2\log\left(\frac{-\epsilon^*_{12}\epsilon_{34}}{16}e^{-\frac{2\pi (t-x)}{\beta}}\right)}.
\ee
Clearly, the OTOC decays polynomially at late time. \\
We have verified this formula numerically for several irrational $\eta$ and in Fig. \ref{fig:N2sr2} we plot the $\mathrm{Im} \ F_{2} (t)$ for $\eta=\sqrt{2}$.

As far as we are aware, this behavior of the OTOC has not been observed before (for all temperatures) and is also our main result for the irrational $\eta$. This finding suggests that, from the perspective of the OTOC, irrational CFTs can be classified as an intermediate step between rational and truly chaotic, holographic CFTs. \\

Let us also remind the reader that this behavior nicely complements our results for the evolution of the second Renyi entropy \cite{Caputa:2017tju}. In that work, we found a new scaling with time of the entropy in the form of double logarithm of time, $\log\log t$. It is tempting to conjecture that, for irrational CFTs, the $\log\log$ evolution of entropies after local operator excitations implies the polynomial decay of the OTOC at late time. We leave a verification of this claim (or a converse statement) as an interesting future problem.

\section{Higher $n$ orbifolds}\label{nHO}
Similarly as in the previous sections, for higher $n$ orbifolds, we distinguish between rational and irrational compactification parameter $\eta$. In principle, for rational $\eta$, the OTOC should again approach to the monodromy constant and, by employing the transformation rules of the higher genus partition functions, one should be able to generalize the RCFT derivation above. In practice, we were not able to perform this algorithm in a simple way and instead we perform a direct computation in the late time limit. Therefore, generalizing our direct approach, we obtain results for both, rational and irrational $\eta$ and verify them numerically. 

The main difficulty here is again to perform the OTO continuation and we briefly summarize the relevant steps. For higher $n$, we have to deal with $\Theta$ functions expressed in terms of $(n-1)\times (n-1)$ matrices that are functions of the two independent, complex cross-ratios $z$ and $\bar{z}$. It turns out that, in order to perform the OTO continuation in $z$, it is sufficient to know how the degenerate hypergeometric functions $_2F_1$ behave under the OTO continuation. This can be done as follows. As shown in \cite{Erdelyi} vol. I, section 2.10, eqs. (14) and (15) , we can rewrite $f_{k/n}(z)$ as
\begin{equation}
f_{k/n}(z)=\f{\sin{\pi \f{k}{n}}}{\pi}\sum_{\m \in \mathbb{Z}} \f{(\f{k}{n})_\m (1-\f{k}{n})_\m}{\m!^2}\PA{\bar{h}_\m-\ln{(1-z)}}(1-z)^{\m},
\end{equation}
where
\begin{equation}
\bar{h}_n=2\psi(1+n)-\psi(k/n+\m)+\psi(1-k/n+\m).
\end{equation}
From this expression, we can show that under the OTO continuation \eqref{OTOCont}, we have
\begin{equation}
f_{k/n}(z) \to f_{k/n}(z)+2 \ii \sin{\PA{\pi \f{k}{n}}} f_{k/n}(1-z).\label{HighernOTO}
\end{equation}
Next, when we consider the late time limit and parametrize the cross-ratios by $(z,\bar{z})\simeq (e^{-\f{\pi}{\d}},  e^{-\f{\pi}{\d^{*}}}) \to (0,0)$, the approximate form of $f_{k/n}$ in the limit $\d \to 0$ becomes
\begin{equation}
f_{k/n}(z) \xrightarrow{\d \to 0}1,~~~~~~~~~~f_{k/n}(1-z) \xrightarrow{\d \to 0}\f{1}{\d}\sin{\PA{\pi \f{k}{n}}}.
\end{equation}
As a result, the expressions for the transformed period matrices become
\begin{eqnarray}
(\tau)_{ij} &\simeq& \f{2}{n} \sum_{k=1}^{n-1} \sin\PA{\pi \f{k}{n}}
\BR{ \f{\sin{\pi\f{k}{n}}}{2 \sin^2 \PA{\pi\f{ k}{n}}-\ii \d} }
\cos \PA{2\pi \f{k}{n}(i-j)},\nonumber\\
(\bar{\tau})_{ij} &\simeq& \f{-2 \ii}{n \d^*} \sum_{k=1}^{n-1} \sin\PA{\pi \f{k}{n}}
\sin\PA{\pi\f{k}{n}}
\cos \PA{2\pi \f{k}{n}(i-j)}\,.
\end{eqnarray}

Analogously to how we proceeded in the $n=2$ case, we first investigate the dominant contribution in the limit $\d^* \to 0$ from the anti-holomorphic part. We find that this contribution is obtained when the following condition is satisfied,
\begin{equation}
\mathbf{n}\eta-\mathbf{m}=\mathbf{0}.\label{Condn}
\end{equation}
For $\eta=\f{p}{p'}$ with $p,p'$ two positive, coprime integers, this condition holds for $\mathbf{m}=p\tilde{\mathbf{m}}, \mathbf{n} = p'\tilde{\mathbf{m}} \ (\tilde{\mathbf{m}} \in \mathbb{Z}^{n-1})$. Hence we can get the following approximate form
\begin{equation}
\begin{aligned}
&\T(0|\text{T}_{mono})\simeq \sum_{\tilde{\mathbf{m}} \in \mathbb{Z}^{n-1}} e^{2 \pi \ii p p'\, \tilde{\mathbf{m}} \cdot \tau \cdot \tilde{\mathbf{m}} }\\
&\simeq \sum_{ \tilde{\mathbf{m}} \in \mathbb{Z}^{n-1}} \EEEX{-\f{2 \pi \ii p p'}{n}\sum_{i,j=1}^{n-1} \tilde{m}_i \tilde{m}_j-\f{1}{2}\f{4\pi  p p'}{n}\d \sum_{i,j,k=1}^{n-1}
\PPA{\f{\cos \PA{2\pi \f{k}{n}(i-j)}}{2\sin^2\PA{\pi \f{k}{n}}}} \tilde{m}_i \tilde{m}_j}.
\end{aligned}
\end{equation}
It is clear from the first term in this expression that the answer should be a function of $pp'$ only and will differ depending if $pp'$ is a multiple of the orbifold number $n$ or not. In fact the same constraint was observed for $n=2$. It is not immediately clear to us what this condition implies physically for the spectrum of the underlying CFT. We have checked that the spectral form factor is also sensitive to this issue but it would be interesting to provide a more physical interpretation for this distinction. Below, we discuss the two cases separately.

If $p p'$ is a multiple of $n$, the first term in the exponent vanishes and, since $\d$ is small, we can approximate the summation by a Gaussian integral. For that, we can introduce a matrix $\ca{A}$
\begin{equation}
\begin{aligned}
(\ca{A})_{ij} \equiv \sum_{k=1}^{n-1}\f{\cos \PA{2\pi \f{k}{n}(i-j)}}{2\sin^2\PA{\pi \f{k}{n}}}=\f{1}{6}(n^2-1)+\abs{i-j}^2-n\abs{i-j},
\end{aligned}
\end{equation}
which has a simple form for the determinant 
\be
\det{\ca{A}}=2^{n-1} n^{n-4}.
\ee
This way, after performing the Gaussian integration, the result for the $\Theta$ function in our limit is written as follows
\begin{equation}
\begin{aligned}
\T(0|\text{T}_{mono})\simeq \PA{\f{n}{2pp'\d}}^{\f{n-1}{2}} (\det{\ca{A}})^{-\f{1}{2}}=\s{\f{n^3}{(4pp'\d)^{n-1}}}.
\end{aligned}
\end{equation}
Moreover, after the OTO continuation, the denominator of $F_n(z,\bar{z})$ can be approximated in the late time limit as
\begin{equation}
\f{1}{\prod_{k=1}^{n-1}\PA{f_{k/n}(z)+2 \ii \sin{\PA{\pi \f{k}{n}}} f_{k/n}(1-z)}f_{k/n}(\bar{z})}
\simeq \f{\pa{-2\d \ii}^{n-1}}{n^2}.\\
\end{equation}
Finally, putting all ingredients together, we obtain the late time constant for $F_n(z,\bar{z})$ in the $n$-th orbifold with $pp'$ being a multiple of $n$ 
\begin{equation}
F_n(z,\bar{z})\simeq \f{n}{(2pp'\ii)^{n-1}}.
\end{equation}

If $pp'$ is not a multiple of $n$, we have to evaluate the complicated summation that appears in function $\T(0|\f{2pp'}{n}\ca{B})$, where
\begin{equation}
\ca{B}_{ij}
\equiv-1+\ii\d\ca{A}_{ij}
=-1+\ii\d\pa{\f{1}{6}(n^2-1)+\abs{i-j}^2-n\abs{i-j}}.
\end{equation}
The analytical approach seems very hard but, after extensive numerical checks, we can estimate this sums as follows. We introduce a ratio $R(pp',n)$
\be
R(pp',n)=\lim_{\delta\to 0}\frac{F^{pp'}_n(z,\bar{z})}{K^{pp'}_n(z,\bar{z})},
\ee
where $F^{pp'}_n(a,\bar{z})$ is just the function $F_n(z,\bar{z})$ for $\eta=p/p'$ and also we defined
\be
K^{pp'}_n(z,\bar{z})\equiv\frac{n}{(2pp'i)^{n-1}},
\ee
which is the result for $pp'$ divisible by $n$. Then for several small values of $pp'$ the results for the ratio are summarized in Table \ref{T1}.
\begin{table}
  \centering
\label{tablefigt}
  \begin{tabular}{|c||c|c|c|c|c|c|c|c|c|c|}
  \hline
  pp' $\setminus$ n & 1 & 2 & 3 & 4  & 5        & 6    & 7    & 8     & 9    & 10 \\ \hline\hline
  1 & 1 & 0 & $-\f{1}{3}$ & $-\f{\ii}{2}$ & $\f{1}{5}$ & 0    & $-\f17$ & $-\f{\ii}{4}$ & $\f19$ & 0 \\ \hline
  2 &   & 1 & $-\f{1}{3}$ & 0         & $\f{1}{5}$ & $-\f13$ & $-\f17$ & $-\f{\ii}{2}$  & $\f19$ & $\f15$ \\ \hline
  3 &   &   & 1         & $\f{\ii}{2}$  & $\f{1}{5}$ & 0    & $-\f17$ & $\f{\ii}{4}$  & $-\f13$ & 0 \\ \hline
  4 &   &   &           & 1         & $\f{1}{5}$ & $-\f13$& $-\f17$ & 0     & $\f19$ & $\f15$\\ \hline
  5 &   &   &           &           &        1 & 0    & $-\f17$ & $-\f{\ii}{4}$ & $\f19$ & 0 \\ \hline
  6 &   &   &           &           &          & 1    & $-\f17$ & $\f{\ii}{2}$  & $-\f13$& $\f15$ \\ \hline
  7 &   &   &           &           &          &      & 1    & $\f{\ii}{4}$  & $\f19$ & 0 \\ \hline
  8 &   &   &           &           &          &      &      & 1     & $\f19$ & $\f15$ \\ \hline
  9 &   &   &           &           &          &      &      &       & 1    & 0 \\ \hline
  10&   &   &           &           &          &      &      &       &      & 1 \\ 
  \hline
\end{tabular}
  \caption{A table of the ratio $R\left(\f{pp'}{n}\right)$ for various $n$ and $pp'$.
  We only presented the result for $n\geq pp'$ because the results opposite case can be obtained
  from the periodicity.}\label{T1}
\end{table}\\
Based on these results, we can conjecture that, for general $n$, $R(pp',n)$ only depends on the ratio $\frac{pp'}{n}$
\be
R(pp',n)=R\left(\f{pp'}{n}\right).
\ee
Clearly, this is consistent with Table \ref{T1}.\\
Moreover, let us reduce the ratio $\frac{pp'}{n}$ into $\f{\ap}{\beta}$ so that $\ap$ and $\beta$ are coprime with each other
\be
\frac{pp'}{n}=\frac{\ap}{\beta},\ \ \ \mbox{such that}\ \ \ g.c.d.(\ap,\beta)=1.
\ee
then, in this general setup, we conjecture that $R\left(\f{pp'}{n}\right)=R\left(\f{\ap}{\beta}\right)$ is given by
\ba
&& \mbox{If}\ \ \  \beta=4\mathbb{Z}:\ \ \  R\left(\f{pp'}{n}\right)=\frac{2}{\beta}\cdot (\ii)^{\ap+2},     \no
&& \mbox{If}\ \ \  \beta=4\mathbb{Z}+1:\ \ \  R\left(\f{pp'}{n}\right)=\frac{1}{\beta},     \no
&& \mbox{If}\ \ \  \beta=4\mathbb{Z}+2:\ \ \  R\left(\f{pp'}{n}\right)=0,     \no
&& \mbox{If}\ \ \  \beta=4\mathbb{Z}+3:\ \ \  R\left(\f{pp'}{n}\right)=-\frac{1}{\beta}.
\ea
This conjecture also reproduces Table \ref{T1} and it would be very interesting to compare it directly with the monodromy constant and the elements of the modular S-matrix \cite{Orbifold}.\\

Let us now discuss higher $n$ orbifolds with irrational $\eta$. In that case, we only have a direct limit computation at our disposal and we carefully extract the limit and compare with numerics. \\
Similarly to $n=2$, if $\eta$ is irrational, the condition \eqref{Condn} holds only if $\mathbf{m}$ and $\mathbf{n}$ vanish. This reduces the higher dimensional $\Theta$ function to identity in this limit and we can approximate function $F_n(z,\bar{z})$ as
\begin{equation}
\begin{aligned}
F_n(z,\bar{z})\simeq \frac{1}{\prod^{n-1}_{k=1}f_{k/n}(z)}.
\end{aligned}
\end{equation}
Now, after taking the OTO continuation and using \eqref{HighernOTO} we derive
\begin{equation}
\begin{aligned}
F_n(z,\bar{z})\simeq\PA{\f{2 \pi \ii }{\ln{z}}}^{n-1} \f{1}{n^2}.
\end{aligned}
\end{equation}
Inserting the explicit form of the cross-ratios yields our final formula for $F_n(z,\bar{z})$ in the $n$-th orbifold for irrational $\eta$
\begin{equation}
\begin{aligned}
F_n(z,\bar{z})\simeq  \f{1}{n^2} \left(\frac{2\pi \ii}{\log\left(-\epsilon^*_{12}\epsilon_{34}e^{-\frac{2\pi(t-x)}{\beta}}\right)}\right)^{n-1}.
\end{aligned}
\end{equation}
This is again the polynomial decay at late time, but interestingly the power of the polynomial increases with $n$. The larger the orbifold number $n$ becomes, the faster the four-point function $F_n$ decays.\\

For convenience, we finish this section with a summary of the results for the OTOCs in general $n$ cyclic orbifolds
\begin{equation}
C_\beta(x,t)\to e^{4\pi \ii \Delta_n}\left\{
    \begin{array}{ll}
     \f{n}{(2pp'\ii)^{n-1}}, &\text{if $\eta=\f{p}{p'} \in \bbb{Q}$ and }  p p' \in n\mathbb{Z},   \\
     \frac{n}{(2pp' \ii)^{n-1}}\cdot R(pp',n),  &\text{if $\eta=\f{p}{p'} \in \bbb{Q}$ and }  p p' \notin n\mathbb{Z}, \\
\PA{\f{2 \pi \ii }{\ln{z}}}^{n-1} \f{1}{n^2},  &\text{if $\eta \notin \bbb{Q}$ }. 
    \end{array}
  \right.\\
\end{equation}
Notice that comparing to the evolution of the second Renyi entropy after local operator excitations \cite{Caputa:2017tju}, OTOCs appear to be more sensitive. Namely, for the Renyi entropies we only distinguished the rational and irrational $\eta$ but OTOCs are very sensitive to $pp'$ already in the rational case. Again, in the spirit of quantum chaos measures, we believe that this fact is related to the sensitivity of the OTOC to the statistics of the spectrum and it would be interesting to verify it directly using for example spectral form factor.

\section{Conclusions}\label{Conclusions}
In this work, we studied the out-of-time-ordered correlators (OTOCs) with twist operators in the cyclic orbifold CFT. We have managed to obtain analytical results for arbitrary $n$ orbifolds that we also confirmed numerically. We verify that for rational compactification parameters the OTOCs behave as expected from the rational CFT. On the other hand, for irrational parameters we find a new, polynomial decay at late time. In Table \ref{T2} we present a summary of our OTOC classification results compared to rational CFTs \cite{Caputa:2016tgt,Gu:2016hoy} as well as large c, holographic CFTs \cite{Roberts:2014ifa,Perlmutter:2016pkf}. In the last column, we show the results for the evolution of the second Renyi entropy \cite{QD,HEE,Caputa:2017tju}. In the table we denoted by $\Delta_{\mathcal{O}}$, the conformal dimension of the operators used in the OTOC and by $d_{\mathcal{O}}$ and $d_{\sigma_n}$ quantum dimension of the local operators. For $n=2$ the late time constant is expressed in terms of the $(S\bar{T}^2 S)_{00}$ as in \eqref{OurRes}. Function $R(pp',n)$ was discussed in the previous section and for $n=2$ we have $R(pp',2)=0$, such that for $pp'\in \text{odd}$ OTOC decays exponentially to zero as in \eqref{OTORCFT}.
\begin{table}[h!]
    \begin{tabular}{ | l | l | l | p{3cm} |}
    \hline
   CFT & Lyapunov & Late Times & $\Delta S^{(2)}_A$ \\ \hline
    RCFT & $\lambda_L=0$ & $M_{00}$ &  $\log d_{\mathcal{O}}$\\ \hline
       $(T^2)^n/\mathbb{Z}_n$, $\eta=\frac{p}{p'}$, $pp'=n\mathbb{Z}$ & $\lambda_L=0$& $e^{4\pi \ii \Delta_n} \f{n}{(2pp'\ii)^{n-1}}$   & $\log d_{\sigma_n}$ \\ \hline
       $(T^2)^n/\mathbb{Z}_n$, $\eta=\frac{p}{p'}$, $pp'\neq n\mathbb{Z}$ & $\lambda_L=0$& $e^{4\pi \ii \Delta_n}\frac{n}{(2pp' \ii)^{n-1}}\cdot R(pp',n)$  & $\log d_{\sigma_n}$ \\ \hline
        $(T^2)^n/\mathbb{Z}_n$ with $\eta\neq\frac{p}{p'}$ & & $\f{e^{4\pi \ii \Delta_n}}{n^2} \frac{(2\pi \ii)^{n-1}}{\left(\log\left(-\epsilon^*_{12}\epsilon_{34}e^{-\frac{2\pi(t-x)}{\beta}}\right)\right)^{n-1}} $ & $(n-1)\log\log(\frac{t}{\epsilon})$  \\ \hline
    Holographic CFT & $\lambda_L=\frac{2\pi}{\beta}$ & $\exp \left(-\frac{2\pi \Delta_{\mathcal{O}} }{\beta}t\right)$ & $2\Delta_{\mathcal{O}}\log\frac{t}{\epsilon}$ \\
    \hline
    \end{tabular}\label{table:Summary}
    \caption{Summary of our results in comparison with the increase in the second Renyi entropy. The blank for the Lyapunov coefficient means that we were not able to draw any conclusions from neither numerics nor analytic arguments.}\label{T2}
\end{table}

At this point, a comment on the Lyapunov coefficients for irrational $\eta$ is in order. In order to see the initial exponential growth of the OTOC in holographic CFTs, the fact that there is a large separation between the collision time $\beta$ and the scrambling time $\sim\beta \log c$ seems very important. In our setup, such a hierarchy is not obvious for all $n$ and indeed our numerical plots (also of the expectation value of the commutator square itself) did not allow us to draw a sufficient conclusions about the Lyapunov exponents form the OTOC. On the other hand, it is possible that in the class of our setup is sub-exponential  as for weak chaos (see e.g. \cite{2015PhRvE..91f2907D}) (at most polynomial, similarly to the polynomial decay at late time) and it is not clear if the OTOC itself is a useful probe in such cases (see also discussion and generalization in \cite{Kukuljan:2017xag}). We hope to return to this problem in the future.

There are several interesting extensions of our work. Firstly, since the correlation function is expressed by the $\Theta(0|\text{T})$ function where the orbifold rank $n$ determines the size of matrix $\text{T}$, it is difficult to understand how the OTOCs behave in the large $n$ limit. This would be crucial to see the Lyapunov exponent in the irrational case and we leave it as an important future work.

Secondly, it would be very interesting to generalize our analysis for the case of symmetric orbifolds of $S_n$. Some results for OTOCs are already known in that setup \cite{Perlmutter:2016pkf} and suggest that the Lyapunov exponent vanishes. It would be interesting to check if the (more)universal OTOCs with twists confirm such behavior for large $n$.

Last but not least, it might be possible to classify 2d CFTs by the way they spread entanglement and quantum information (hence their potential to describe black holes). The evolution of Renyi entropies and OTOCs make this more concrete and exploring them in other interesting 2d CFTs should fill the gaps in this exciting project.

\section*{ Acknowledgements}
We would like to thank Gabor Sarosi, Ida Zadeh, Alvaro Veliz-Osorio and Tokiro Numasawa for discussions and comments on the manuscript.
TT and PC are supported by the Simons Foundation through the ``It from Qubit'' collaboration. KW is supported by JSPS fellowship. TT is supported by JSPS Grant-in-Aid for Scientific Research (A) No.16H02182 and World Premier International Research Center Initiative (WPI Initiative) from the Japan Ministry of Education, Culture, Sports, Science and Technology (MEXT). 

\appendix

\section{Cross-ratios}\label{CROTO}
\begin{figure}[htbp]
 \begin{center}
  \includegraphics[clip,width=7.5cm]{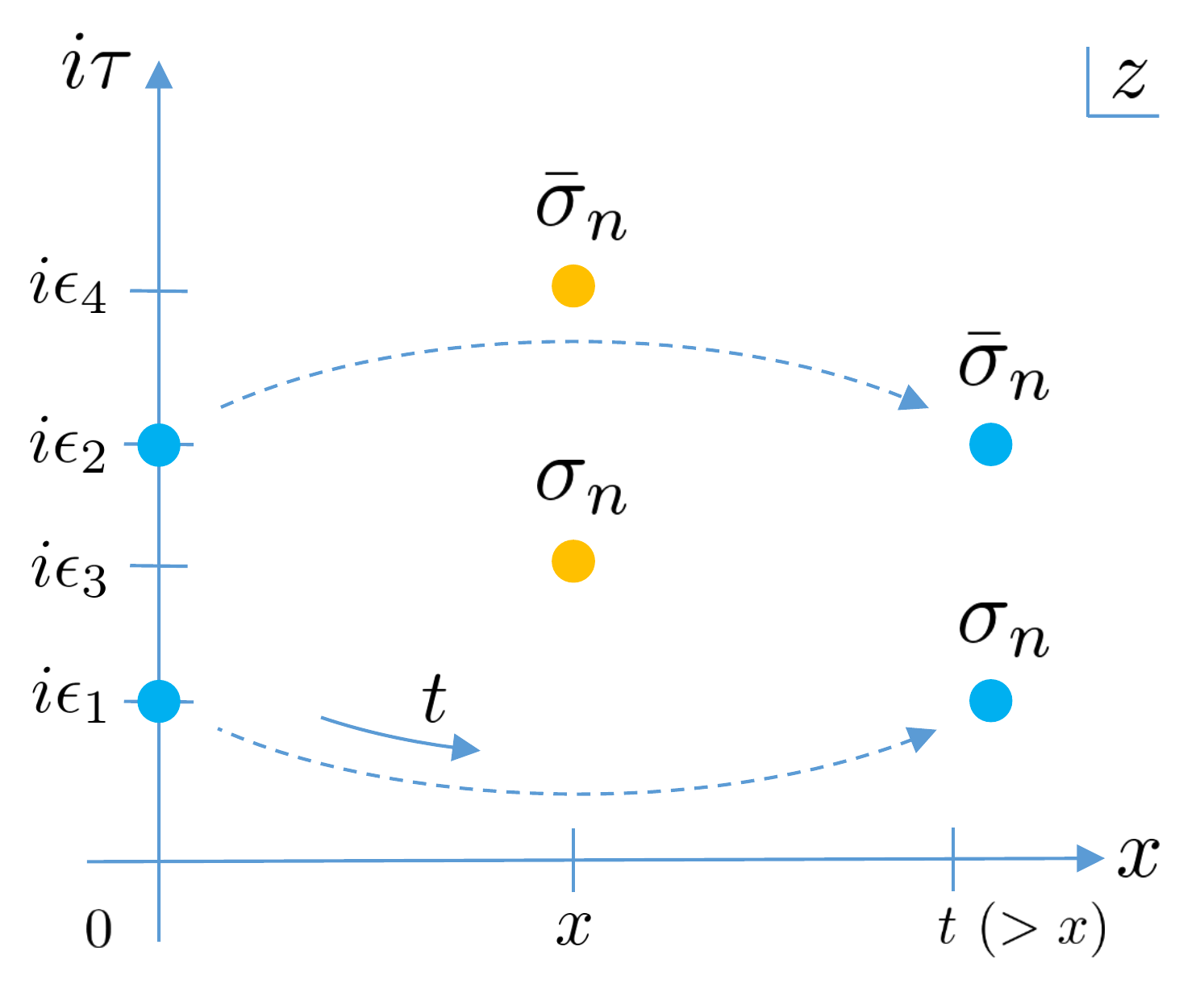}
 \end{center}
 \caption{Ordering of the twist operators in the OTOC. dashed lines denote the "trajectory" of the two operators with the progress of time.}\label{fig:OPS}
\end{figure}

The out-of-time-ordered correlators can be obtained from the standard four-point functions by analytic continuation. This is done by ordering operators along imaginary time with $\epsilon$'s and continuing to the real time (see the main text). The conformal cross ratios on the cylinder become
\bea
z&=&\frac{z_{12}z_{34}}{z_{13}z_{24}}=\frac{\sinh\left(\frac{\pi \ii(\epsilon_1-\epsilon_2)}{\beta}\right)\sinh\left(\frac{\pi \ii(\epsilon_3-\epsilon_4)}{\beta}\right)}{\sinh\left(\frac{\pi (t-x+\ii\epsilon_1- \ii\epsilon_3)}{\beta}\right)\sinh\left(\frac{\pi (t-x+ \ii\epsilon_2-\ii\epsilon_4)}{\beta}\right)},\nn
\bar{z}&=&\frac{\bar{z}_{12}\bar{z}_{34}}{\bar{z}_{13}\bar{z}_{24}}=\frac{\sinh\left(\frac{\pi \ii(\epsilon_1-\epsilon_2)}{\beta}\right)\sinh\left(\frac{\pi \ii(\epsilon_3-\epsilon_4)}{\beta}\right)}{\sinh\left(\frac{\pi (t+x+\ii\epsilon_1-\ii\epsilon_3)}{\beta}\right)\sinh\left(\frac{\pi (t+x+\ii\epsilon_2-\ii\epsilon_4)}{\beta}\right)}.
\eea
The $\epsilon$'s are ordered such that $\epsilon_1<\epsilon_3<\epsilon_2<\epsilon_4$, and as time progresses, the operators change their relative position on the complex plane (see Fig. \ref{fig:OPS}). This is equivalent to the OTO continuation.
Then, at late time we can approximate the denominators such that the cross ratios approach to $0$ as
\bea
z\simeq -e^{-\frac{2\pi (t-x)}{\beta}}\epsilon^*_{12}\epsilon_{34},\qquad
\bar{z}\simeq -e^{-\frac{2\pi (t+x)}{\beta}}\epsilon^*_{12}\epsilon_{34},
\eea
where $\epsilon_{ij}=\ii\left(e^{\frac{2\pi \ii}{\beta}\epsilon_i}-e^{\frac{2\pi \ii}{\beta}\epsilon_j}\right)$.

\end{document}